\documentclass[a4paper,12pt]{article}
\pdfoutput=1 

\usepackage{jheppub} 

\usepackage[T1]{fontenc} 

\usepackage{caption}
\usepackage{cancel}
\usepackage{enumitem}

\usepackage{amsmath,amscd}
\usepackage{amsfonts}
\usepackage{amssymb,scalerel,stackengine}
\usepackage{amsmath,amsfonts,amssymb,amsthm,bbm,bm}
\usepackage{relsize}
\usepackage{empheq}
\usepackage{graphicx,shortvrb}
\usepackage{comment}
\usepackage[utf8]{inputenc}
\numberwithin{equation}{section}

\hypersetup{
    colorlinks,%
    citecolor=blue,%
    filecolor=blue,%
    linkcolor=blue,%
   urlcolor=blue,
   linktoc=page
}

\def\bz{{\bar z}}
\def\mn{{\mu\nu}}
\def\lm{{\ell m}}
\def\l{\lambda}

\def\de{\delta}

\def\pd{\partial}

\def\cd{\nabla}
\def\eps{\epsilon}
\def\veps{\varepsilon}
\def\cL{\mathcal{L}}
\def\cM{\mathcal{M}}

\def\cF{\mathcal{F}}
\def\cG{\mathcal{G}}
\def\cP{\mathcal{P}}
\def\cQ{\mathcal{Q}}
\def\cV{\mathcal{V}}

\def\cO{\mathcal{O}}
\def\cI{\mathcal{I}}

\def\tl{\tilde}
\def\avg{\mathcal{R}}

\title{Gravitational breathing memory and dual symmetries}

\author{Ali Seraj}

\affiliation[]{Centre for Gravitational Waves, Universit\'{e} Libre de Bruxelles\\International Solvay Institutes, CP 231, B-1050 Brussels, Belgium}

\emailAdd{aseraj@ulb.ac.be}

\abstract{\noindent Brans-Dicke theory contains an additional propagating mode which causes homogeneous expansion and contraction of test bodies in transverse directions. This ``breathing'' mode is associated with novel memory effects in addition to those of general relativity. Standard tensor mode memories are related to a symmetry principle: they are determined by the balance equations corresponding to the BMS symmetries. In this paper, we show that the leading and subleading breathing memory effects are determined by the balance equations associated with the leading and ``overleading'' asymptotic symmetries of a dual formulation of the scalar field in terms of a two-form gauge field. The memory effect causes a transition in the vacuum of the dual gauge theory. These results highlight the significance of dual charges and the physical role of overleading asymptotic symmetries.}

\begin{document} 
\maketitle
\flushbottom

\section{Introduction}

The gravitational wave (GW) is not purely an oscillatory waveform, but it includes a non-oscillatory part that builds up over time and leads to a final offset in the waveform. This is the so-called GW ``memory'' effect. When such a waveform passes through a set of test masses which are initially at rest, the GW memory causes a permanent displacement in their relative distances. This ``displacement'' memory effect is known since the 70's \cite{Zeldovich:1974gvh}. However, the early understanding constituted only part of the memory, nowadays known as the linear or ordinary part. It was later realized by Christodoulou \cite{Christodoulou:1991cr} and independently by Blanchet and Damour \cite{Blanchet:1992br} that the radiation of massless fields, including GWs, leads to another contribution to the memory which is called the non-linear or null memory. The construction of \cite{Christodoulou:1991cr} was based on an asymptotic analysis of Einstein equations in asymptotically flat spacetimes. Another outcome of the asymptotic analysis is the appearance of BMS symmetries in asymptotically flat spacetimes \cite{Bondi:1962px,Sachs:1962wk} as an infinite dimensional extension of the Poincar\'e algebra. However, the interplay between the GW memory and BMS symmetries was uncovered only recently \cite{Strominger:2014pwa}. Indeed, Strominger and collaborators discovered the so-called IR triangle, which displayed the correspondence between three low energy effects in general relativity (GR), namely the BMS group, displacement memory effect, and Weinberg's soft graviton theorem \cite{Strominger:2017zoo}. The correspondence also holds at subleading level where the subleading soft graviton theorem is related to an extension of BMS algebra by super Lorentz symmetries \cite{Barnich:2010eb,Campiglia:2014yka,Campiglia:2015yka}. The corresponding memories are called the spin \cite{Pasterski:2015tva} and center of mass memories \cite{Nichols:2018qac}.

Similar construction applies to gauge theory \cite{Strominger:2017zoo}, where the large gauge symmetries play the role of BMS symmetries and the associated memory is a kick effect \cite{Bieri:2013hqa,Pasterski:2015zua}. However, a natural puzzle arises for a massless scalar QFT in Minkowski spacetime: which symmetry would be responsible for the soft factorization in this theory? In \cite{Campiglia:2017dpg} a set of charges were build such that the correspondence holds. It was shown in \cite{Campiglia:2018see,Francia:2018jtb} that these charges can be formulated as the Noether charges associated to the asymptotic symmetries of a dual formulation of the theory in terms of two form fields. Asymptotic symmetries and charges of $p$ form gauge theories in four and higher dimensions are studied in  \cite{Afshar:2018apx} (see \cite{Esmaeili:2020eua} for a review). Indeed, defining charges in a theory by resorting to the symmetries of a dual formulation of the theory is not restricted to massless scalars, but also includes magnetic type charges in Maxwell theory \cite{Hosseinzadeh:2018dkh,Freidel:2018fsk,Henneaux:2020nxi} and gravity \cite{Godazgar:2018qpq,Kol:2019nkc,Godazgar:2019dkh,Godazgar:2020kqd,Oliveri:2020xls}. 

What is interesting about scalar tensor theories and particularly Brans-Dicke (BD) theory is that the non-minimal coupling between the scalar and the metric leads to an additional \textit{gravitational} degree of freedom. This transverse scalar mode, leads to homogeneous expansion and contraction of test bodies in the transverse plane of the GW propagation, thus also called the ``breathing'' mode. More formally, considering a congruence of geodesics subject to GWs, the $+,\times$ modes lead to shearing effects, while the scalar mode causes an expansion in the evolution of the congruence.  Accordingly, BD theory has an additional breathing memory on top of the usual memories present in GR. The breathing memory was first found by Lang  \cite{Lang:2013fna,Lang:2014osa} and some of its observational aspects are investigated in \cite{Du:2016hww,Koyama:2020vfc}. Recently, Tahura et al. \cite{Tahura:2020vsa} found a ``subleading'' breathing memory effect in BD theory by considering the congruence of timelike geodesics  with initial relative velocity.

The BMS symmetries and their balance equations in BD theory were studied in detail in \cite{Hou:2020wbo,Tahura:2020vsa,Hou:2020tnd}. These balance equations determine the memory effects in tensorial modes of the GW. However, the additional breathing memories in BD theory, have no obvious symmetry counterpart, as the theory has no additional gauge symmetry.  This led \cite{Tahura:2020vsa} to conclude that the new memories have no symmetry interpretation\footnote{Also, the viewpoint in \cite{Hou:2020tnd} was to associate the breathing memory to the angular momentum balance equation. However, reserving that equation to constraint the spin and center of mass memories \cite{Nichols:2017rqr,Nichols:2018qac}, one needs a new symmetry to describe the breathing memory effect.}. However, in this work, we show that the missing symmetries reside in the dual formulation of the theory. We show that the duality between scalar and two form fields continues to hold in the gravitational context of Brans-Dicke theory. Using this fact, and inspired by \cite{Campiglia:2018see}, we study the asymptotic symmetries of a dual formulation of BD theory and show that they lead to a tower of charges and balance equations. These charges contain exactly what is needed to describe the gravitational breathing memories.

The leading breathing memory of \cite{Lang:2013fna}, the scalar soft theorem of \cite{Campiglia:2017dpg}, and the leading scalar dual charges constitute the corners of the an IR triangle for the scalar field in BD theory. Moreover, the subleading breathing memory \cite{Tahura:2020vsa} and the corresponding charges that we find in this paper suggest that there is a subleading soft theorem for the scalar field in BD theory. We expect that this subleading soft theorem is the one discussed in \cite{DiVecchia:2015jaq}. Also the subsubleading soft scalar theorem found in \cite{DiVecchia:2015jaq}, suggest that one can extend the IR triangle even one step further. We leave these issues for a later study.

The paper is organized as follows: In section \ref{sec: memory in BD}, we review Brans-Dicke theory and study its asymptotic structure at null infinity. We then discuss inertial frames and  derive the breathing memory. We largely follow \cite{Tahura:2020vsa}. However, we manage to simplify their main result for the memory in a way that facilitates the study of corresponding symmetries. In section \ref{sec: memory and symmetry}, we introduce the dual formulation of BD theory by starting from the Einstein formulation. We then study the asymptotic behavior of the dual theory, derive its symmetries and the corresponding charges using a symplectic structure build from the covariant phase space approach. In appendix \ref{appendix null dyad}, we provide some details on null congruences and inertial frames in BD theory.

\section{Memory effects in Brans-Dicke theory}\label{sec: memory in BD}
The action of Brans-Dicke theory in the Jordan frame is given by 
\begin{align}
    S=\frac{1}{16\pi}\int d^{4} x \sqrt{-g}\left[\lambda R-\frac{\omega}{\lambda} g^{\mu \nu} \partial_{\mu} \lambda\partial_{\nu} \lambda\right]\,.
\end{align}
Physical distances are measured by the metric $g_\mn$ and matter fields are supposed to be minimally coupled to this metric. The vacuum field equations then read
\begin{subequations}
\begin{align}\label{eom Jorda}
G_{\mu \nu} &=\frac{1}{\lambda}\left(T_{\mu \nu}^{(\lambda)}+\nabla_{\mu} \nabla_{\nu} \lambda-g_{\mu \nu} \square \lambda\right)\,, \\
\square \lambda &=0 \,,
\end{align}
\end{subequations}
where 
\begin{align}
    T_{\mu \nu}^{(\lambda)} &=\frac{\omega}{ \lambda}\left(\nabla_{\mu} \lambda \nabla_{\nu} \lambda-\frac{1}{2} g_{\mu \nu} \nabla^{\alpha} \lambda \nabla_{\alpha} \lambda\right)\,.
\end{align}
We do not include matter fields in the action, as our analysis will be in the far zone where matter fields are not present. Adding electromagnetic field is possible and will not alter the field equations for the scalar $\l$ as the electromagnetic stress tensor is traceless. Also the corrections to the metric are mostly subleading except certain contributions which can be added a posteriori.
\subsection{Asymptotic analysis in flat spacetime}
In order to discuss the memory effect non-perturbatively, it is necessary to perform an asymptotic analysis of the field equations, far from the source.
A convenient setup for this analysis is the Bondi-Sachs formalism, in which the fields are expanded in negative powers of the distance, and the equations are solved order by order. A detailed analysis of the Brans-Dicke theory in the Bondi-Sachs formalism was performed recently in \cite{Tahura:2020vsa,Hou:2020tnd}, both in Einstein and Jordan frames.The starting point is to define the coordinate system $(u,r,x^A)$  consisting of the retarded time $u$, radial distance $r$ and angular coordinates $x^A$ on the sphere. The coordinate system is fixed by the Bondi gauge conditions
\begin{align}\label{Bondi gauge}
    g_{rr}=g_{rA}=0, \qquad  \partial_{r} \operatorname{det}\left(r^{-2} g_{A B}\right)=0\,.
\end{align}
The gauge conditions \eqref{Bondi gauge} imply that  $u$ is null everywhere, $r$ is the areal distance, and $x^A$ are asymptotically constant along outgoing null rays. In this gauge, the metric takes the general form
\begin{align}\label{metric Bondi}
    ds^2&=e^{2\beta}({U}du^2-2du\,dr)+r^2 \gamma_{AB}(dx^A-U^Adu)(dx^B-U^Bdu)
\end{align}
with $\sqrt{-g}=e^{2\beta}(\det g_{AB})^{1/2}$ and the inverse
\begin{align}
g^{\mu v}=\left(\begin{array}{ccc}
0 & -e^{-2\beta} & 0 \\
-e^{-2\beta} & -Ue^{-2\beta} & -e^{-2\beta}U^{B} \\
0 & -e^{-2\beta}U^{A} & \frac{1}{r^2}\gamma^{A B}
\end{array}\right)\,.
\end{align}
For this metric to be asymptotically flat, we impose the following falloff behavior
\begin{align}\label{BC}
    \gamma_{A B} &=q_{A B}(x^A)+\cO (1/r), \qquad \l=\l_0(u,x^A) +\cO(1/r)
\end{align}
where the leading metric $q_{AB}$ on the ``celestial'' sphere is assumed to be time independent. As conventional in Bondi-Sachs formalism, we use $q_{AB}$ to define the covariant derivative $D_A$ and the Levi-Civita tensor $\eps_{AB}$ on the sphere. We also use this fixed metric to lower and raise indices on the sphere.
Equations \eqref{Bondi gauge} and \eqref{BC} imply that at any radius,
\begin{align}\label{det g_AB}
    \det (g_{AB})=r^4\det ( q_{AB})
\end{align}
Following Bondi, we may impose as a boundary condition that $q_{AB}=\mathrm{diag}(1,\sin^2\theta)$ is the metric of the round sphere . The other option is to let $q_{AB}(x^A)$ be an arbitrary time independent metric on the sphere\footnote{It is even possible to allow time dependence for $q_{AB}$. This is consistent with Penrose conformal compactification of asymptotically flat spacetimes. However, this choice complicates the asymptotic behavior  \cite{Barnich:2010eb}. Removing the time dependence of $q_{AB}$ is related to choosing divergence free conformal completion of spacetime, see \cite{Ashtekar:2014zsa}.} with the only constraint that $\oint d^2x \sqrt{q}=4\pi$. The second option allows to extend the BMS algebra with super-Lorentz symmetries \cite{Barnich:2010eb,Campiglia:2014yka,Compere:2018ylh}. In both cases, the area of any coordinate spheres is $4\pi r^2$ for any asymptotically flat geometry within the class \eqref{metric Bondi}.
The boundary conditions \eqref{BC} are enough to determine asymptotically the rest of the fields at null infinity up to integration constants \cite{Tahura:2020vsa,Hou:2020tnd}. Let us perform an asymptotic expansion for the four quantities appearing in the metric \eqref{metric Bondi} and the scalar field as 
\begin{align}\label{asymptotic expansion metric Jordan}
\begin{split}
U &= \mathring{U} +2m\,r^{-1}+\cO(r^{-2}) \\
\beta &=\mathring{\beta}\,r^{-1}+\cO(r^{-2}) \\
\gamma_{A B} &=q_{A B}+ C_{A B}\,r^{-1}+D_{A B}\,r^{-2}+\cO(r^{-3}) \\
U^{A} &=\mathring{U}^{A}r^{-2}+\cO(r^{-3})\\
\l&=\l_0+\l_1\,r^{-1}+\l_2\,r^{-2}+\cO(r^{-3})\,.
\end{split}
\end{align}
The scalar field equation implies that at leading order
\begin{align}\label{eom phi0}
\pd_u\l_0=0, \qquad D^2\l_0=0\,.    
\end{align}
where $D^2\equiv D_AD^A$ is the Laplacian on the sphere. Therefore, assuming smoothness, $\l_0$ is a free constant on the sphere. The next order $\l_1(u,x^A)$ is free and includes the radiative part of the scalar field at null infinity. At next order we get
\begin{align}\label{balance eq scalar}
    \pd_u\l_2&=-\frac{1}{2}D^2\l_1\,,
\end{align}
which will be important to describe the subleading displacement memory in the scalar sector. On the other hand, Einstein equations reveal that $q_{AB}, C_{AB}$ are free data, and that no terms  can appear in \eqref{asymptotic expansion metric Jordan} more leading than those indicated. Einstein equations also determine some other coefficients algebraically in terms of the free data
\begin{align}
\mathring U= -\frac12\, R[q]-\frac{\dot\lambda_1}{\lambda_0}\,,\qquad
\mathring\beta=&-\frac{\lambda_{1}}{2 \lambda_{0} }\,\qquad \mathring U^{A}=-\frac{1}{2 }\left(D_{B} C^{A B}-\frac{D^{A} \lambda_{1}}{\lambda_{0}}\right)\,,
\end{align}
where an overdot refers to a derivative with respect to $u$. The Einstein equations also imply the following balance equation for the modified mass aspect $\cM$
\begin{align}\label{eom mass aspect}
    \partial_{u} \mathcal{M}=-\frac{1}{8}\Big( \dot{C}_{A B} \dot{C}^{A B}+\frac{2\Omega^2}{\lambda_{0}^{2}}\,\dot \lambda_{1}^{2}\Big)+\frac{1}{4} \Big(D_{A} D_{B} \dot{C}^{A B}+\frac{1}{\lambda_{0}} D^2 \dot\lambda_{1}\Big)\,,
\end{align}
where
\begin{align}
 \mathcal{M}\left(u, x^{A}\right)\equiv m-\frac{1}{4 \lambda_{0}^{2}} \lambda_{1} \partial_{u} \lambda_{1}\,,\qquad\Omega\equiv \sqrt{2 \omega+3}  \,.
\end{align}
The modified mass aspect is defined such that its monopole moment (which measures the energy as we will see) is monotonically decreasing in time.

\subsection{Memory effects}
Gravitational memory effects are permanent changes in the metric of spacetime after the passage of gravitational waves. In BD theory, there is an additional memory effect with respect to GR, known as the breathing memory. To observe these memories, we study the effect of the GW on a set of test masses far from the source. More precisely, consider a local congruence of timelike geodesics in a region far from the source. Construct an orthonormal frame $e_{\hat \mu}=e_{\hat \mu}^{\;\;\,\nu}\pd_\nu$ adapted to the transverse plane, i.e. $ e_{\hat A}$ basis vectors are tangent to the transverse plane, $ e_{\hat r}$ is along the GW propagation and $ e_{\hat 0}$ is tangent to a timelike geodesic, representing the origin of the inertial frame. As we explain in appendix \ref{appendix null dyad} , at leading order in $\cO(1/{ r})$,
\begin{align}\label{inertial frame}
 e_{\hat 0}=\partial_{  u}-\frac{1}{2 \lambda_{0}} \dot{\lambda}_{1} {\partial}_{  r},\qquad  e_{\hat r}=\partial_{  u}-\Big(1+\frac{1}{2 \lambda_{0}} \dot{\lambda}_{1}\Big) \partial_{  r},\qquad  e_{\hat A}=\frac{1}{  r} \mathrm{e}_{\hat A}\,.
\end{align}
The dyad $\mathrm{e}_{\hat A}$ is obtained from $q_{AB}$ by $q_{AB}\mathrm{e}^{\;\;A}_{\hat A}\mathrm{e}^{\;\;B}_{\hat B}=\de_{\hat A \hat B}$. One can also define a dual basis $e^{\hat \mu}=e^{\hat \mu}_{\;\;\,\nu}d x^\nu$ where $e^{\hat \mu}_{\;\;\,\nu}\equiv\eta^{\hat \mu \hat \alpha}\,g_{\nu\beta}\,e_{\hat \alpha}^{\;\;\,\beta}$.
We are particularly interested in the evolution of timelike geodesics in the transverse plane, which can be described by studying the shape tensor $\cd_{(A}\smash{ e^{\hat 0}}_{B)}$ which turns out to take the asymptotic form
\begin{align}\label{Theta def}
   \cd_{(A}\smash{ e^{\hat 0}}_{B)}=-\frac{r}{2}\dot\Theta_{AB}+\cO(1)\,,\qquad\Theta_{AB}\equiv C_{AB}-\frac{\lambda_{1}}{\lambda_{0}}  q_{AB}\,.
\end{align}
This indicates that the congruence of geodesics experience a shearing encoded in the Bondi shear $C_{AB}$, and an expansion  determined by the scalar field
\begin{align}\label{Theta trace}
    \Theta\equiv q^{AB}\Theta_{AB}=-2\frac{\lambda_{1}}{\lambda_{0}}\,,
\end{align}
as a result of Einstein equations.
The combination $\Theta_{AB}$ will show up frequently in the following. Therefore, BD theory has an additional transverse degree of freedom, leading to homogeneous expansion and contraction of test bodies. This is why it is called the breathing mode.

\paragraph{Geodesic deviation.} To make the above more precise, we study the evolution of the distance between neighboring geodesics in the inertial frame, given by the geodesic deviation equation
\begin{align}
    \ddot{X}_{\hat{i}}=-R_{\hat{0}\hat{i}\hat{0} \hat{j} } X^{\hat{j}}+\cO(  r^{-2})\,,
\end{align}
where $X_{\hat{i}}$ is the position of a test mass with respect to a reference one, considered as the origin, and $\hat i,\hat j$ refer to spatial indices. The leading gravitational effect appears at order $1/r$ in the transverse components of the Riemann tensor $R_{\hat{0}\hat{A}\hat{0} \hat{B} }$. Thus, the relevant equation is 
\begin{align}\label{geodesic deviation}
    \ddot{X}_{\hat{A}}=-R_{\hat{0}\hat{A}\hat{0} \hat{B} } X^{\hat{B}}+\cO(  r^{-2})\,.
\end{align}
The Riemann component takes the following  form over the solutions 
\begin{align}\label{Riemann-Bondi}
    R_{\hat{0} \hat{A} \hat{0} \hat{B}}
    &=-\frac{1}{2   r}\ddot\Theta_{\hat{A} \hat{B}}\,.
\end{align}
where $\Theta_{AB}$ was introduced in \eqref{Theta def}. Now suppose that the gravitational field is radiating only in the time interval between  an initial time $u_0$ and a final time $u_f$. Outside this time interval there is no gravitational wave and hence $C_{AB}, \l_1$ are constants. 
Using \eqref{Riemann-Bondi} in \eqref{geodesic deviation} and integration over time, one finds
\begin{align}\label{test mass position}
{X}_{\hat{A}}(u)=X_{\hat{A}}^{(0)}+(u-u_0)\dot{X}_{\hat{A}}^{(0)}+\frac{1}{2r} \int_{u_0}^{u}du'\int_{u_0}^{u'}du''  \Theta_{\hat{A} \hat{B}} {X}^{\hat{B}}\,,
\end{align}
where $X_{\hat{A}}^{(0)},\dot{X}_{\hat{A}}^{(0)}$ are the initial position and velocity of the body in the transverse plane. The first two terms simply describe the free motion in flat spacetime, while the last term is the correction due to the gravitational wave. 

\begin{figure}[h]
    \centering
    \captionsetup{width=.8\linewidth}
    \includegraphics[width=0.4\columnwidth]{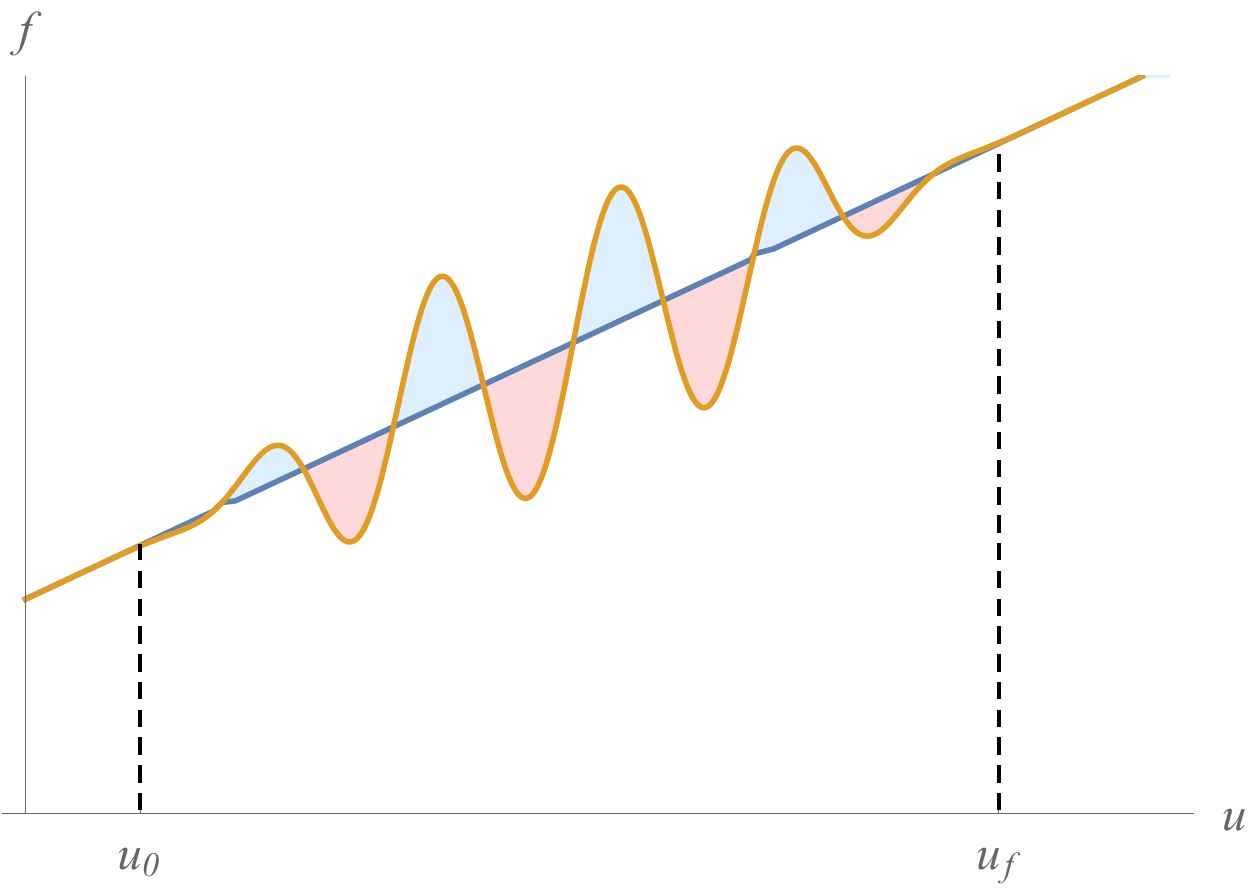}
    \caption{For an arbitrary function $f$, the operator $\Delta$ captures the total change in the function between the initial and final time. On the other hand, $\avg$ captures the difference between the area under $f$ and its ``trapezoidal'' average $\overline f$.  }
    \label{fig 1}
\end{figure}

\paragraph{Memory effects.} We are interested in permanent effects after the passage of GW, e.g. in $\Delta X_{\hat{A}}=X_{\hat{A}}(u_f)-X_{\hat{A}}(u_0)$. This was computed in \cite{Tahura:2020vsa}. Their result can be simplified and written in a suggestive form after introducing the following notation. Define the operation $\avg$ as
\begin{align}
    \avg f\equiv \int_{  u_{0}}^{  u_{f}} d   u \big(f-\overline{f}\big)\,,\qquad \overline{f}\equiv \frac12\big(f(u_0)+f(u_f)\big)\,.
\end{align}
In figure \ref{fig 1}, we depict the effect of this operator. The total displacement of the test mass in the interval $\Delta u=u_f-u_0$ is given by (\textit{cf.} eq. (4.17) of \cite{Tahura:2020vsa})
\begin{equation}\label{memory-full-BD}
   \boxed{ \Delta X_{\hat{A}}= \Delta u \dot{X}_{\hat{A}}^{(0)}+\frac{1}{2  r}\Big\lbrace X_{0}^{\hat{B}}\,\Delta \Theta_{\hat{A} \hat{B}} -2\dot{X}_{0}^{\hat{B}}\,\avg \Theta_{\hat{A} \hat{B}}\Big\rbrace\,.}
\end{equation}

Let us elaborate on this result. The overall effect of the passage of the gravitational wave on the test mass is encoded into two memory effects in the final state of the system at $u\geq u_f$: the \textit{displacement} memory and the \textit{kick} memory. Explicitly
\begin{align}
    {X}_{\hat{A}}(u)=X_{\hat{A}}(u_f)+(u-u_f)\dot{X}_{\hat{A}}(u_f)\,,\qquad u\geq u_f\,,
\end{align}
where the coefficients are given by 
\begin{align}
    X_{\hat{A}}(u_f)=X_{\hat{A}}^{(0)}+(u_f-u_0)\dot{X}_{\hat{A}}^{(0)}+\mkern-15mu\underbrace{d_{\hat{A}}/r}_{\substack{\text{displacement} \\ \text{memory}}}\mkern-10mu,\quad \dot{X}_{\hat{A}}(u_f)=\dot{X}_{\hat{A}}^{(0)}+\mkern-5mu\underbrace{v_{\hat{A}}/r}_{\substack{\text{kick} \\ \text{memory}}}\,.
\end{align}
The displacement and kick memories are given by 
\begin{subequations}
\begin{align}
 d_A&= \frac{1}{2} X_{0}^{\hat{B}}\,\Delta \Theta_{\hat{A} \hat{B}} -\dot{X}_{0}^{\hat{B}}\,\avg \Theta_{\hat{A} \hat{B}}\,,\label{displacement memory} \\
 \qquad v_A&= -\frac12 \dot X_{0}^{\hat{B}}\,\Delta \Theta_{\hat{A} \hat{B}} \,.\label{kick memory}
\end{align}
\end{subequations}
The displacement memory involves a differential and an integral effect, given by $\Delta,\avg$ , which are referred to as \textit{leading} and \textit{subleading} displacement memory effects respectively.

\subsection{Memory and balance equations}
The balance equation \eqref{eom mass aspect} can be rearranged as 
\begin{align}\label{balance eq 1}
    \frac{1}{4} \partial_{u} D^{A} D^{B}\left(C_{A B}+\frac{\lambda_{1}}{\lambda_{0}} \delta_{A B}\right)=\partial_{u} \mathcal{M}+\frac{1}{8} \dot{C}_{A B} \dot{C}^{A B}+(3+2 \omega) \frac{1}{4 \lambda_{0}^{2}}\left(\partial_{u} \lambda_{1}\right)^{2}\,.
\end{align}
Multiplying this with an arbitrary function $\veps(x^A)$ and integrating over the sphere and the time interval, we find
\begin{align}\label{memory constraint 1}
    \frac{1}{4} \int_{S^{2}} \varepsilon D^{A} D^{B}\Delta C_{AB}=\Delta \cP_{\varepsilon}+\mathcal{F}_{\varepsilon}\,,
\end{align}
where 
\begin{align}\label{supermomentum def}
\cP_{\varepsilon}=\int_{S^{2}} \varepsilon\Big(\mathcal{M}-\frac{1}{4 \lambda_{0}}  \lambda_{1}\Big)
\end{align}
is the \textit{supermomentum} charge with parameter $\veps$ and its associated quadratic flux through the boundary is
\begin{align}
    \cF_\veps=\frac{1}{8} \int_{\Delta\cI} \veps\Big(\dot{C}_{A B} \dot{C}^{A B}+(3+2 \omega) \frac{1}{4 \lambda_{0}^{2}}\left(\partial_{u} \lambda_{1}\right)^{2}\Big)\,.
\end{align}
Throughout the paper, we use the shorthand notation
\begin{align}
	\int_{\Delta \cI} \equiv \lim_{\rho \to\infty} \int_{u_0}^{u_f} du\int_{S^2},\qquad \int_{S^2}\equiv \int_{S^2} d^2x \sqrt{q}
\end{align}

We will justify in the next section that eq.\eqref{supermomentum def} reduces to the standard expression for the supermomenta in the Einstein frame. Equation \eqref{memory constraint 1} determines the tensorial (shearing) part of the memory. However, the memory given by eq. \eqref{memory-full-BD}, is characterized by the tensor $\Theta_{AB}$, which also contains a memory in its trace $\Theta$, called the breathing memory. Therefore, it is desirable to treat $\Delta \Theta$ in a similar manner, i.e. to write it in terms of a balance equation. To this end, we use the result \eqref{Theta trace} to arrive at 
\begin{align}\label{memory flux scalar}
    \frac18\int_{S^{2}}\varepsilon D^2\Delta\Theta=\Delta Q_{\varepsilon}
\end{align}
where the quantity
\begin{align}\label{charge scalar}
Q_{\varepsilon}=-\frac{1}{4\lambda_{0}} \int_{S^{2}} \lambda_{1} D^2 \varepsilon    
\end{align}
can be thought of as a charge associated with the scalar field. We will justify this in the next section by showing that it is the Noether charge associated to the asymptotic symmetries of the dual formulation of the scalar field in BD theory. Subtracting eqs. \eqref{memory constraint 1} and \eqref{memory flux scalar}, we find 
\begin{align}\label{memory constraint 2}
    \frac{1}{4} \int_{S^{2}} \varepsilon D^{A} D^{B}\Delta\Theta_{AB}=\Delta \cP_{\varepsilon}+\mathcal{F}_{\varepsilon}+\Delta Q_{\varepsilon}\,.
\end{align}
In the next subsection, we show how the subleading displacement memory is fixed by another set of charges.
\subsection{Subleading memory}
The subleading part of the displacement memory is given by the second term in \eqref{displacement memory}. It is again sourced by tensorial and scalar modes. The former is related to the balance equations for the super Lorentz charges, as described in detail in \cite{Tahura:2020vsa,Compere:2019gft}. Let us focus here on the latter, i.e. the scalar contribution to the subleading memory. This part is characterized  by
\begin{align}\label{memory constraint 3}
D^2\avg\Theta&=\int_{  u_{0}}^{  u_{f}} d   u   D^2\big(\Theta-\overline{\Theta}\big)=-\frac{2}{\l_0}\int_{  u_{0}}^{  u_{f}} d   u  D^2\big(\l_1-\overline{\l_1}\big)
\end{align}
Using the balance equation \eqref{balance eq scalar} for the scalar field, we can write the above expression smeared with a function $D^2\veps$ as 
\begin{align}\label{memory constraint subleading}
    -\frac18\int_{S^{2}} D^2\varepsilon D^2\avg\Theta&=\Delta\cQ_\veps+\Delta u\,\overline{Q}_{D^2\veps}
\end{align}
where the new charge is defined as 
\begin{align}\label{charge scalar subleading}
    \cQ_\veps=-\frac{1}{2\lambda_{0}} \int_{S^{2}} \lambda_{2}  D^2\varepsilon    
\end{align}
We will show in the next section that this additional charge $\cQ_\veps$ is the Noether charge associated to \textit{overleading} dual symmetries of the scalar field.
\subsection{Inversion of the constraints}
The constraints \eqref{memory flux scalar} and \eqref{memory constraint subleading} can be inverted to find constraints over the memory effects appearing in \eqref{memory-full-BD}. This can be done either by expansion in terms of spherical harmonics, or more formally through Green's functions. Defining $Q_{\ell m}\equiv Q_{\veps =Y_{\ell m}^\ast}$ and $\Delta \Theta=\sum_{\ell,m}\Delta \Theta^{\lm} Y_{\lm}$, we find from \eqref{memory flux scalar} that
\begin{align}
     \Delta \Theta^{\lm}=-\frac{8}{\ell(\ell+1)}\Delta Q_{\lm},\qquad \ell\geq 1
\end{align}
Therefore apart form the monopole of the breathing mode, all the multipole moments are given in terms of the fluxes of charges defined in \eqref{charge scalar}. One can make the picture more coherent by defining a monopole charge
\begin{align}\label{monopole charge}
    Q_{0}\equiv-\frac{1}{2\pi\l_0}\int_{S^2} \l_1
\end{align}
The above results can be written in a more compact form by inverting the Laplacian i.e. a Green function $G_{y}(x)=G(x,y)$ such that 
\begin{align}
    D^2 G(x,y)=\frac{1}{\sqrt{q}}\de^2 (x-y)-\frac{1}{4\pi}\,,\qquad  x,y\in S^2.
\end{align}
The constant term is necessary so that  both sides of the equation integrate to zero  over the sphere.
The Green function is explicitly given by
\begin{align}
G(x,y)=\frac{1}{4\pi}\log \sin^2\frac{|x-y|}{2}+\text{const.}    
\end{align}
where $|x-y|$ is the polar separation between the points $x^A,y^A$ on the sphere. 
Using this in \eqref{memory flux scalar}, and implementing the monopole charge \eqref{monopole charge}, we find 
\begin{align}
    \Delta \Theta(y)&=8\Big(\Delta q+\Delta Q[\veps(x)=G(x,y)]\Big)
\end{align}
where $y$ is a given point on the sphere and $x$ is a dummy variable integrated over in the charge integral. While the definition of the monopole charge $Q_0$ might look ad hoc at this stage, we will see in the next section that it is indeed a Noether charge with completely different origin from the higher multipole charges \eqref{charge scalar}.
Finally, we can invert the constraint \eqref{memory constraint subleading} using a similar method, leading to 
\begin{align}
     \avg \Theta^{\lm}=\frac{8}{\ell^2(\ell+1)^2}\Big(\Delta \cQ_{\lm}-\ell(\ell+1)\Delta u\,\overline{Q}_{\lm}\Big) ,\qquad \ell\geq 1
\end{align}
It is not clear for us whether the monopole moment of $\avg \Theta$ can be described using a flux equation or not. 
\subsection{Asymptotics in Einstein frame}
By a redefinition of the fields as $\tl g_\mn=\frac{\l}{\l_0}{g}_\mn, \l=\exp(\phi/\Omega)$ with $\Omega=\sqrt{2\omega+3}$, the action of Brans-Dicke theory is transformed, up to surface terms, to that of GR with a minimally coupled scalar $\phi$
\begin{align}\label{action Einstein frame}
    S=\l_0\int d^{4} x \sqrt{-\tilde{g}}\Big[\tilde{R}-\frac{1}{2} \tilde{g}^{\mn}\pd_\mu\phi\pd_\nu\phi\Big]
\end{align}
The metric $\tl{g}_\mn$ as defined above satisfies the gauge conditions \eqref{Bondi gauge} except the determinant condition, since $\sqrt{\det \tl{g}_{AB}}=\frac{\l}{\l_0}\sqrt{\det {g}_{AB}}=\frac{\l}{\l_0}r^2\sqrt{q}$. This is because the radial coordinate is the areal distance with respect to ${g}_\mn$ and not $\tl{g}_\mn$. It is still possible to perform a simple redefinition of the radial coordinate as
\begin{align}\label{rho vs r}
    \rho&=r\sqrt{\frac{\l}{\l_0}}=r+\frac{\l_1}{2\l_0}+\cO(1/r)
\end{align}
so that the metric $\tl g_\mn(u,\rho,x^A)$ is in Bondi gauge. This implies that the conformal map between the two frames should be accompanied by a coordinate transformation. Explicitly, denoting $\tl x^\mu=(u,\rho,x^A)$, $ x^\mu=(u,r,x^A)$, then
\begin{align}\label{conformal map}
    \tl g_\mn(\tl x)=\frac{\l(x)}{\l_0}\frac{\pd x^\alpha}{\pd \tl x^\mu}\frac{\pd x^\beta}{\pd \tl x^\nu}{g}_{\alpha\beta}(x),\qquad \l(x)=\l_0 \exp\Big(\frac{\phi(\tl x)-\phi_0}{\Omega}\Big)
\end{align}
Using \eqref{conformal map} and the results of the previous section, one can obtain the asymptotic behavior of the fields in the Einstein frame. 
In particular, assuming an asymptotic expansion for the scalar as  $\phi=\sum_{n=0}^\infty \phi_n(u,x^A)\rho^{-n}$, we can find the relation between the scalars in the two frames. Note that
\begin{align}
    e^{(\phi-\phi_0)/\Omega}&=1+\frac{1}{ \rho\Omega}\phi_1( u,x^A)+\frac{1}{ \rho^2\Omega}\big(\phi_2+\frac{\phi_1^2}{2\Omega}\big)+\cdots
\end{align}
However, after replacing $\rho$ with $r$ perturbatively using \eqref{rho vs r} we find that
\begin{align}\label{lambda vs phi}
     \frac{\l_1}{\l_0}=\frac{\phi_1}{\Omega},\qquad \frac{\l_2}{\l_0}=\frac{\phi_2}{\Omega}
\end{align}
and a balance equation holds for the subleading component of the scalar
\begin{align}\label{balance scalar Einstein}
    \dot \phi_2&=-\frac{1}{2}D^2\phi_1
\end{align}
On the other hand, the metric being in Bondi gauge takes the form 
\begin{align}\label{metric Bondi Einstein}
    ds^2&=e^{2b}\big({V}du^2-2du\,d\rho\big)+\rho^2 h_{AB}(dx^A-V^Adu)(dx^B-V^Bdu)
\end{align}
\begin{align}
h_{AB}&= {\tl q}_{AB}+{\rho}^{-1} {\tl C}_{AB}+\cO(\rho^{-2})\,,  &&b=\mathring b\,\rho^{-2}+\cO(\rho^{-3})\nonumber\\
{V}^{A} &=\mathring{V}^A\rho^{-2}  +\cO(\rho^{-3})\,,  &&{V}=\mathring V-2 M\rho^{-1}+\cO\left({\rho}^{-2}\right)\,,
\end{align}
where the relationship between these asymptotic coefficients and those of Jordan frame \eqref{asymptotic expansion metric Jordan} is obtained as
\begin{align}
\tl q_{AB}&= q_{AB}\,\qquad \tl C_{AB}=C_{AB}\,,\qquad
    \mathring{b}=\mathring{\beta}- \frac18 \phi_{1}^{2}\,,\qquad  \mathring{V}^A=\mathring U^A\,,\qquad \mathring V=-\frac{R[q]}{2}
\end{align}
Note that there is no radial dependence in the above quantities, e.g. $\tl C_{AB}(u,x^A)=C_{AB}(u,x^A)$. Therefore, in the following, we will always replace $\tl q_{AB},\tl C_{AB}$ in favor of the ones without tilde. Finally, the relationship between the mass aspects in the two frames is 
\begin{align}\label{mass aspects relationship}
    M=m-\frac{\phi_1\dot\phi_1}{4\Omega^2}+\frac{1}{2\Omega}\dot\phi_2=\cM-\frac{1}{4\l_0} D^2\l_1\,.
\end{align}
Given that in the Einstein frame, the theory is Einstein GR plus a minimally coupled scalar, it is well known that the supermomenta take the form $\cP_\veps=\oint \veps M$. Using \eqref{mass aspects relationship}, we see that this coincides with \eqref{supermomentum def} in the Jordan frame.
\section{The interplay between memory and symmetry}\label{sec: memory and symmetry}
In previous section, it was shown that the description of the breathing memory motivates two sets of charges given in eqs. \eqref{charge scalar},\eqref{charge scalar subleading}. Given that BD theory has no additional local symmetry other than the diffeomorphisms, it is natural to ask which symmetries these charges correspond to? In \cite{Campiglia:2018see}, this question was partly addressed in a non-gravitational context, where it was shown that the leading charges \eqref{charge scalar} correspond to the asymptotic symmetries of a dual formulation of the scalar in terms of two-form gauge fields. Here we extend their results to the gravitational context. Moreover, we show that the subleading charges \eqref{charge scalar subleading} are also Noether charges associated to a set of ``overleading'' symmetries of the gauge field.

\subsection{Dual formulation of the scalar}
\paragraph{Introduction.} Let us consider a theory of two form field $B=\frac{1}{2}B_\mn dx^\mu dx^\nu$ with the Lagrangian $\cL=-\frac{1}{2} {H} \wedge * {H}$, where $H=dB$ is the corresponding field strength\footnote{For a rank $k$ form in $n$ dimensions, we define $\alpha=\frac{1}{k!}\alpha_{i_1\cdots i_k}dx^1\wedge\cdots\wedge dx^k$ and we have
\begin{align*}
 (\ast \alpha)_{i_{k+1}\cdots i_n}&=\frac{1}{k!}\eps_{i_1\cdots i_n}\alpha^{i_1\cdots i_k}\\
 (d\alpha)_{i_1\cdots i_{k+1}}&=(k+1)\pd_{[i_k+1}\alpha_{i_1\cdots i_k]}
\end{align*}
where $\eps$ is the Levi-Civita \textit{tensor} including $\sqrt{|g|}$.}. The field equation and the Bianchi identity read
\begin{align}\label{2form eom}
  d\ast H=0 \, ,\qquad dH=0\,,
\end{align}
which imply respectively that
\begin{align}\label{2fom sol}
    H=\ast d \phi\,,\qquad \Box\phi=d\ast d\phi=0\,,
\end{align}
for some scalar field $\phi$. Replacing the first equality back into the Lagrangian, we find that $\cL=-\frac{1}{2} d \phi \wedge * d \phi$, which coincides with that of a free scalar field. In this sense, the two form theory is dual to a free scalar theory. Note that the role of field equation and Bianchi identity is exchanged when going to the dual picture.
This construction is independent of the metric of spacetime and therefore can be used in a gravitational theory as well. The only consistency requirement imposed by the Bianchi identity is that the field equation for the scalar field has no source term. This is the case for Brans-Dicke theory even in the presence of electromagnetic matter fields. Therefore, in this section we propose to consider the following dual action instead of \eqref{action Einstein frame}:
\begin{align}\label{action EH 2form}
S=\int d^{4} \tl x \sqrt{-{\tl g}}\left[\tl{R}-\frac{1}{12} H_{\mn\alpha} H^{\mn\alpha}\right]   \,,
\end{align}
where the tilde reminds that we are working in the Einstein frame, where the coordinates are $\tl x^\mu=(u,\rho,x^A)$ and the metric is given by \eqref{metric Bondi Einstein}. For simplicity, we do not put tilde over the gauge fields, which should not lead to any confusion.
\subsection{Asymptotic behavior of the gauge field}
We simplify the analysis by imposing the following gauge conditions on the gauge field
\begin{align}
B_{u\rho}=B_{uA}=0\,,\qquad D^A B_{\rho A}=0\,.
\end{align}
The latter condition is primarily imposed on an initial hypersurface $u=const$. However, the field equations extend it to allover the spacetime\footnote{I thank Erfan Esmaeili for reminding me this point.}. As in the previous section, we use the leading metric $q_{AB}$ to raise and lower indices on the sphere, and to define the connection $D_A$.
Now consider the following asymptotic expansion for the remaining components
\begin{align}\label{B falloff}
\begin{split}
{B}_{\rho A} &=\sum_{n=-\infty}^{N-1} \rho^n B_{\rho A}^{(n)}(u,x^A), \qquad {B}_{A B} =\sum_{n=-\infty}^{N} \rho^{n} B_{A B}^{(n)}(u,x^A) 
\end{split}
\end{align}
where $N$ is a positive integer that we will come back to soon. Note that contrary to the expansion of the scalar field $\phi$, here positive $n$ refers to the coefficient of a positive power of $\rho $. This asymmetric notation simplifies the results in the following. To find the asymptotic behavior of the gauge field, we need to solve eq.\eqref{2fom sol} for the gauge field $B$. Note that 
\begin{align}\label{H falloff-B}
\begin{aligned}
{H}_{u \rho A} &=\pd_{u}B_{\rho A}\,,\qquad {H}_{u A B} =\pd_{u}B_{AB}\,,\qquad {H}_{\rho  A B} =\pd_\rho  B_{ A B}+ 2\pd_{[A}B_{B]\rho }
\end{aligned}
\end{align}
while the right hand side of \eqref{2fom sol} is given by 
\begin{subequations}\label{H falloff-scalar}
\begin{align}
(\tl\ast d\phi)_{u \rho  A} &=-\eps_{AB}(e^{2b}h^{BC}\pd_C\phi-\rho^2U^B\pd_\rho\phi)=-\eps_{A}^{\;\;\;B} \pd_B\frac{\phi_1}{\rho}+\cO(\rho^{-2} )\\
(\tl\ast d\phi)_{u A B} &=\rho ^2\epsilon_{AB}\big(\dot \phi+V\pd_\rho \phi+V^A\pd_A\phi\big)=\rho\,\eps_{AB}\, \dot\phi_1+\cO(1) \\
(\tl\ast d\phi)_{\rho  A B} &=-\rho ^2\epsilon_{AB}\pd_\rho \phi=\eps_{AB}(\phi_1+2\frac{\phi_2}{\rho })+\cO(1/\rho ^2)\label{H rho AB 2}
\end{align}
\end{subequations}
To compare the two form theory and the scalar theory more easily, we note that the gauge condition $D^A B_{\rho A}=0$ allows to write $B_{\rho A}$ in terms of one scalar field\footnote{According to the Hodge decomposition, we can write $B_{\rho A}$ into exact and co-exact pieces. The gauge condition $D^AB_{\rho A}=0$ kills the exact part.}. On the other hand $B_{AB}$ being a two form on the sphere is necessarily proportional to the Levi-Civita tensor, so we can write\footnote{In defining $\sigma$, we have made explicit a derivative with respect to $\rho$ which simplifies our later computations. This does not restrict the solutions once we add a logarithmic term in the expansion of $\sigma$.} 
\begin{align}\label{Bab decomposed}
    B_{\rho A}=\eps_{A}^{\;\;\;B}\pd_B\pd_\rho\sigma, \qquad B_{AB}= \eps_{AB}\psi
\end{align}
Using \eqref{Bab decomposed} in \eqref{H rho AB 2}, we get
\begin{align}\label{H rho AB}
    {H}_{\rho  A B} =\pd_\rho  B_{ A B}+ 2\pd_{[A}B_{B]\rho }=\eps_{AB}\pd_\rho \left(\psi+D^2\sigma\right)
\end{align}
where we have used the fact that in two dimensions, antisymmetrization of two indices can be trade off with the Levi-Civita tensor
\begin{align*}
    2u_{[A}v_{B]}=(\de_A^{\;\;C}\de_B^{\;\;D}-\de_B^{\;\;C}\de_A^{\;\;D})u_Cv_D=\eps_{AB}\eps^{CD}u_Cv_D
\end{align*}
Comparing \eqref{H rho AB} and the third equation in \eqref{H falloff-scalar}, we find  
\begin{align}
    \pd_\rho \big(\psi+D^2\sigma\big)+\rho ^2\pd_\rho \phi&=0\,,
\end{align}
Using the asymptotic expansion 
with an asymptotic expansion 
\begin{align}\label{asymptotic expansion psi}
    \sigma &=\sum_{n=-\infty}^{N} \rho ^n \sigma_{n}+\bar\sigma \log\rho, \qquad \psi =\sum_{n=-\infty}^{N} \rho ^n \psi_{n}\,.
\end{align}
we find
\begin{subequations}
\begin{align}\label{psi n expanded}
    \psi_1&=\phi_1-D^2\sigma_1\,,\\
    \psi_{n}&=-D^2\sigma_{n}\,,\qquad n\geq 2\,,
\end{align}
\end{subequations}
Also equating the first and second equations in \eqref{H falloff-B} and \eqref{H falloff-scalar}, we find that 
\begin{align}
    \bar\sigma=-\int du \phi_1\,,\qquad\text{and}\qquad \pd_u\sigma_n=0\,,\quad  n\geq 0
\end{align}
This suffices for our analysis of symmetries and charges in the next section.
\subsection{Phase space structure}
The asymptotic solutions discussed in the previous section can be given a canonical structure. This is most conveniently obtained by using the covariant phase space approach \cite{Ashtekar:1981bq,Wald:1999wa}.  The covariant phase space is defined as the collection of asymptotic solutions accompanied by a symplectic structure. Focusing on the two form field, we find through a standard procedure that the symplectic current as a spacetime three form is given by 
\begin{align}
\omega_B=\delta \ast H\wedge \delta {B}=\de d\phi\wedge \delta {B}
\end{align}
where the $\wedge$ denotes antisymmetrization of the phase space variations, combined with the spacetime wedge product\footnote{More explicitly, when the symplectic current is contracted with two variations $\de_1,\de_2$, we have $\Omega_B(\de_1,\de_2)=\frac12\,\de_1 d\phi\wedge \delta_2 {B}-(1\leftrightarrow 2)$ where now the symbol $\wedge$ simply refers to the spacetime wedge product.}.
The symplectic structure of the radiative phase space is obtained by integrating the symplectic current over a worldtube at large constant radius between the initial time $u_0$ when the system is nonradiative to $u_f$ at which the system again goes back to a nonradiative configuration. The relevant component is $\omega^B_{uAB}=\frac12\,\de \dot\phi\wedge \delta B_{AB}$. Using \eqref{Bab decomposed} and taking the integral, we find
\begin{align}\label{Radiative symplectic structure}
    \Omega_B&=\int_{\Delta \cI} \de  \dot \phi\wedge \de\psi
\end{align}
With the general asymptotic behavior for $B_{AB}$ implied by \eqref{asymptotic expansion psi}, the symplectic structure $\Omega_B$ diverges as $\rho^{N-1}$. The divergence can be removed by imposing the boundary condition $B_{AB}=\rho B^{(1)}_{AB}+\cO(1)$, as done in \cite{Campiglia:2018see}, leading to the following symplectic structure of the \textit{radiative phase space}
\begin{align}\label{symplectic structure radiative}
\Omega_{(1)}&=\int_{\Delta \cI} \de  \dot \phi_1\wedge \de \psi_1= \int_{\Delta \cI} \de  \dot \phi_1\wedge \de (\phi_1-D^2\sigma_1)\,.
\end{align}
A more heuristic approach, which we will follow in this work is to define a the symplectic structure as the finite part of the bare symplectic form \eqref{Radiative symplectic structure}\footnote{While it may be possible to formulate this more rigorously through a renormalization process, we will not study this issue here.}. The regularized symplectic structure is found  by making an asymptotic expansion of all fields and keeping the $\rho$ independent piece of the bare symplectic structure \eqref{Radiative symplectic structure}. We then find
\begin{align}
    \Omega_{\text{reg}}=\sum_{n=1}^N \Omega_{(n)},\qquad \Omega_{(n)}=\int_{\Delta\cI}\, \de  \dot \phi_n\wedge \de \psi_n
\end{align}
It turns out that this approach is more informative than the other one, as it allows to define a tower of \textit{overleading} symmetries, their charges and balance equations which in turn allow to discuss all memory effects coherently. Overleading symmetries have shown up in different fashions in the literature, including multipole symmetries in gauge theory and gravity \cite{Seraj:2016jxi,Seraj:2017rzw,Compere:2017wrj,Kutluk:2019ghr}, as well as those inspired by subleading soft theorems \cite{Campiglia:2016efb,Campiglia:2018dyi,Laddha:2017vfh,Conde:2016csj} and adiabatic modes \cite{Mirbabayi:2016xvc,Hamada:2018vrw}.

Note that the symplectic structure of the scalar theory and its dual two form theory are not the same. The symplectic structure of the scalar field is given by $\Omega_\phi=\int_{\Delta \cI} \de  \dot \phi_1\wedge \de \phi_1$. Comparing this with \eqref{Radiative symplectic structure}, we see that they differ by 
\begin{align}\label{Omega 1}
    \Omega_{\text{reg}}-\Omega_\phi&=-\sum_{n=1}^N\int_{\Delta \cI} \de  \dot \phi_n\wedge \de D^2\sigma_n=-\sum_{n=1}^N\int_{S^2} \de  \Delta \phi_n\wedge \de D^2\sigma_n
\end{align}
where we used the fact that $\sigma_n$ is time independent for $n\geq 1$. The full symplectic structure contains additional non-radiative \textit{edge} modes that lead to  nontrivial symmetries and charges, which in turn explain the memory effects.  

Adding the standard contribution from the GR sector \cite{Ashtekar:1981bq}, the full symplectic structure of the theory is given by 
\begin{subequations}
\begin{align}\label{symplectic structure full}
    \Omega&=\int_{\Delta \cI}\left(\de  \dot C^{AB}\wedge \de C_{AB}+\de \dot\psi_1\wedge \delta \psi_1\right)-\sum_{n=2}^N \int_{S^2} \de \Delta \phi_n\wedge \de D^2\sigma_n\\
    &=\int_{\Delta \cI}\left(\de  \dot C^{AB}\wedge \de C_{AB}+\de \dot\phi_1\wedge \delta \phi_1\right)-\sum_{n=1}^N \int_{S^2} \de \Delta \phi_n\wedge \de D^2\sigma_n
\end{align}
\end{subequations}
In the last line, The first integral denotes the radiative phase space of gravitational degrees of freedom, while the second integral represents the phase space of edge modes whose conjugate momenta are related to the gravitational memories as we will discuss. 
\subsection{Symmetries and balance equations}
The theory is \eqref{action EH 2form} is invariant under transformations
\begin{align}
    B\to B+\alpha, \qquad d\alpha=0
\end{align}
Moreover, $\alpha$ should have the same asymptotic behavior as in eq.\eqref{B falloff} and should preserve the temporal gauge we have imposed. 
The solution is given by $\alpha=-d\beta$ with 
\begin{align}
    \beta=\beta_A dx^A\,, \qquad \beta_A=\eps_A^{\;\;\;B}\pd_B\,\mu(\rho ,x^A)\,.
\end{align}
In terms of the fields defined in \eqref{Bab decomposed}, this corresponds to the shift
\begin{align}\label{homogeneous solutions}
\sigma\to \sigma- \mu(\rho ,x^A)\,,\qquad     \psi\to \psi +D^2\mu (\rho ,x^A)
\end{align}
In other words, by expanding the symmetry parameter as $\mu=\sum_{n=-\infty}^{N} \rho ^{n} \veps_{n}(x^A)$, we find a tower of symmetry transformations 
\begin{align}\label{shift symmetry}
    \sigma_n\to \sigma_n -\veps_n(x^A)\,\qquad \psi_n\to \psi_n+D^2\veps_n(x^A)\,.
\end{align}
Whether these are redundant pure gauge transformations or physical asymptotic symmetries is determined by the symplectic structure of the theory. If the contraction of the symplectic structure with a variation $\de_\veps$ is zero, then this variation is a degeneracy of the symplectic form and hence quotiented out in the physical phase space. However, if $\Omega(\de,\de_\veps)\neq0$ for some $\de$ tangent to the phase space, then $\veps$ represents an asymptotic symmetry and is associated with a nontrivial charge on the physical phase space. Noting \eqref{symplectic structure full}, we see that $\veps_n$ is nontrivial for $ n\geq 1$.

\paragraph{Tower of Charges and fluxes.} By contracting the symplectic structure with a symmetry transformation, we can define a flux associated to that symmetry, i.e. $\de\cF_\veps=\Omega(\de,\de_\veps)$. Using \eqref{symplectic structure full}, and considering the symmetry transformation \eqref{shift symmetry},  
\begin{align}
    \cF_n[\veps]=\int_{S^2} \Delta\phi_n D^2\veps\,,\qquad n\geq 1
\end{align}
The flux can be written as the change in a Noether charge associated to the symmetry. In other words, we have the flux-balance equation $\cF_n[\veps]=-\Delta Q_n[\veps]$ where
\begin{align}\label{tower of charges}
    Q_n[\veps]=-\int_{S^2}\phi_n D^2\veps\,,\qquad n\geq 1
\end{align}
In summary, we have a tower of charges labeled by $n$ and at each level, we have infinitely many charges parameterized by an arbitrary function $\veps_n(x^A)$ on the sphere.
Also it is important to note that the charges $Q_1[\veps]$ does not contain the monopole charge as the monopole is in the kernel of the Laplacian appearing in the definition of the charge \eqref{tower of charges}. However, looking at the second term in the symplectic structure \eqref{symplectic structure full}, we see that a constant shift $\phi\to\phi+c$ is an independent symmetry of the symplectic form and leads to an additional charge
\begin{align}\label{monopole charge Einstein}
    q=-\int_{S^2}\phi_1
\end{align}
This charge corresponds to the shift symmetry of the original scalar theory. Therefore, the full symmetry algebra is the union of the symmetries of the dual frames. This is similar to electric and magnetic charges in Maxwell theory; while the (asymptotic) electric charges appear naturally as Noether charges corresponding to the large gauge symmetries in the standard description of the theory, magnetic charges play the same role in the dual description of the theory. Remarkably, both set of charges are physical observables in either of the descriptions and are given by local expressions of the fields. 

Let us focus on the first sets of charges $q,\,Q_1[\veps],\, Q_2[\veps]$. Note that using eq.\eqref{lambda vs phi}, we get
\begin{align}
    q=4\Omega Q_0\,,\qquad Q_1[\veps]=4\Omega  Q_\veps\,,\qquad Q_2[\veps]=2\Omega \cQ_\veps
\end{align}
Therefore, up to trivial normalization constants, these are the charges whose balance equations determine the breathing memory effects in Brans-Dicke theory, as was shown in the previous section, eqs. \eqref{monopole charge}, \eqref{charge scalar} and \eqref{charge scalar subleading}. 

\subsection{Memory as vacuum transition}
There is a second aspect of the relationship between memory effects and BMS symmetries. Suppose that the spacetime is non-radiative before and after a burst of gravitational wave lasting between the time interval $(u_0,u_f)$. From the perspective of the radiative phase space given by the symplectic structure \eqref{symplectic structure radiative}, the system is in  vacuum state before $u_0$ and after $u_f$. However, these vacua are not identical precisely due to the memory effect.
Therefore, instead of a single vacuum, the radiative phase space possesses a ``space of vacua''. The space of vacua is a subspace of the covariant phase space on which the pull back of the symplectic form vanishes. From \eqref{symplectic structure radiative}, we see that this subspace is given by time independent functions $\psi_1(x^A)$. The gravitational memory effect causes a transition between an initial vacuum $\psi_1^{(0)}$ and a final one $\psi_1^{(f)}$. Now we show that the vacuum transition can be modeled by a symmetry transformation. To see this, note that by construction,  the charges introduced in the previous section satisfy
\begin{subequations}
\begin{alignat}{4}
    \big\{Q_1[\veps],\psi_1\big\}&=D^2\veps\,, \qquad&\big\{Q_1[\veps],\sigma_1\big\}&=-\veps\,\\
    \big\{q,\psi_1\big\}&=1\,, &\big\{q,\sigma_1\big\}&=0
\end{alignat}
\end{subequations}

Therefore we can find a constant $c$ and a function $\veps$ such that 
\begin{align}
     \psi_1^{(f)}&=\psi_1^{(0)}+\big\{c\,q+Q_1[\veps],\psi_1^{(0)}\big\}
\end{align}
Also note that according to \eqref{psi n expanded}, $\phi_1=\psi_1+D^2\sigma_1$ and therefore
\begin{align}
     \big\{Q_1[\veps],\phi_1\big\}&=0
\end{align}
Therefore $\phi_1$ is gauge invariant which \textit{explains} why there is no gauge symmetry in the scalar field formulation.
\section{Discussion}
In this work, we discussed gravitational memory effects in Brans-Dicke theory. This can be considered as a means to test general relativity, as a nonvanishing breathing memory signals deviation from general relativity \cite{Du:2016hww}. It is desirable to study various permanent effects due to the scalar mode and to compute the breathing memory for particular dynamical sources, for example how scalarization processes are encoded in the memory. The memory effects in more realistic scalar-tensor theories are even more interesting, as the scalar memory can be significantly enhanced as a result of the screening mechanism \cite{Koyama:2020vfc}. 

At a theoretical level, we showed that the breathing memory effect in BD theory can be explained through a symmetry principle by taking into account ``dual symmetries'', i.e. the asymptotic symmetries of a dual formulation of the scalar field in terms of two-form gauge fields. Our construction relied on the fact that the scalar field is massless and free. This is not the case in the presence of generic matter fields, or in more general scalar-tensor theories. This suggests that there exists a more general way to define dual symmetries, e.g. through an asymptotic notion of the duality \cite{Freidel:2018fsk}.

Our construction revealed a tower of asymptotic charges given in eq. \eqref{tower of charges}. At $n=2$, a remaining issue is that the subleading memory can contain monopole mode, which is not captured by the charges that we obtained. For more subleading charges $n> 2$, it is important to see if they are related to new low energy effects. The subsubleading soft dilaton factorization of \cite{DiVecchia:2015jaq}, valid at tree level, suggests that this is the case. 

We obtained the tower of charges through a regularized symplectic structure obtained through a finite part prescription. It is desirable to find this result in a more systematic fashion, e.g. by adding suitable counterterms in the action.

\paragraph{Acknowledgments}
I thank Erfan Esmaeili and David Nichols for useful discussions. The author is funded by the European Union's Horizon 2020 research and innovation program under the Marie Sklodowska-Curie grant agreement No 801505.

\appendix

\section{Null congruences in Bondi gauge}\label{appendix null dyad}
To compare the asymptotic behavior in BD and GR, it would be useful to study the behavior of the congruence of future directed outgoing and ingoing null dyad
\begin{align}\label{null dyad}
    \ell=\pd_r\,,\qquad    n=n_\mu dx^\mu=e^{-2\beta}(\pd_u+\frac{U}{2}\pd_r+U^A\pd_A)\,,
\end{align}
The latter can be written as a dual 1-form as $n=\frac{U}{2}du-dr$ from which it is obvious that $\ell\cdot n=-1$. The ingoing null vector takes the asymptotic form $n=\pd_u+\frac{\mathring U}{2}\pd_r+\cO(1/r)$, so that in Jordan frame
\begin{align}
    n=\pd_u-\frac12(1+\frac{\dot\l_1}{\l_0})\pd_r+\cO(1/r)
\end{align}
Remind that a null congruence with tangent vector $k$ can be described in terms of expansion, shear and twist  
\begin{align}
    \cd_Ak_B=\frac12\theta g_{AB}+\sigma_{AB}+\omega_{AB}
\end{align}
with 
\begin{align}
    \theta=g^{AB}\cd_Ak_B,\qquad \sigma_{AB}=\cd_{(A}k_{B)}-\frac12\theta g_{AB},\qquad \omega_{AB}=\cd_{[A}k_{B]}
\end{align}
Both the congruences of \eqref{null dyad} are twist-free. For the outgoing congruence of $\ell$, the expansion and shear coincide asymptotically with that of GR
\begin{align}
     \cd_{(A}\ell_{B)}=rq_{AB}+\frac{1}{2}C_{AB}+\cO(1/r)
\end{align}
while for the ingoing vector $n$, there is a dynamical mode in its expansion 
\begin{align}
    \cd_{(A}n_{B)}=-\frac{r}{2}\Big(\dot C_{AB}+q_{AB}\mathring{U}\Big)+\cO(1)
\end{align}
since $\mathring{U}=-(\frac12R[q]+\frac{\l_1}{\l_0})$. From the above null vectors, we can construct unit timelike and spacelike vectors 
\begin{align}
    \hat e_0=n+\ell/2\,,\qquad \hat e_r=n-\ell/2
\end{align}
reproducing \eqref{inertial frame}. An interesting property of Bondi gauge in GR is that the Bondi frame asymptotically reduces to an inertial frame, as can be seen by setting $\lambda=0$ in \eqref{inertial frame}. However, this is not anymore the case in BD theory. A test mass at rest in the inertial frame has four-velocity $$v^\mu\pd_\mu=\hat e_0=\partial_{  u}-\frac{1}{2 \lambda_{0}} \dot{\lambda}_{1} {\partial}_{  r}.$$ Therefore, this worldline is described in the Bondi coordinates as radially oscillating. However, this coordinate effect disappears when studying geodesic \textit{deviation}. It is also possible to modify the Bondi gauge such that  the coordinate and inertial frames coincide asymptotically. One has to replace \eqref{det g_AB} by the modified determinant condition
\begin{align}\label{modified det condition}
    \det (g_{AB})=r^4(\frac{\l}{\l_0})^2\det ( q_{AB})
\end{align}
This implies that the Bondi tensor $C_{AB}$ will have a trace (expansion) part as well. Imposing \eqref{modified det condition} might look strange, but it is natural from the perspective that the transverse field $C_{AB}$ contains all the propagating degrees of freedom, consisting of the shearing $+,\times$ polarization modes, as well the scalar breathing mode. It is also equivalent to imposing that $r$ is the areal distance but with respect to the metric in the Einstein frame. This approach was followed in \cite{Hou:2020tnd}.

\bibliography{Refs}
\end{document}